\documentclass[aps,twocolumn,reprint]{revtex4-1}

\draft 

\usepackage[T2A]{fontenc}
\usepackage[utf8]{inputenc}
\usepackage{wrapfig}
\usepackage[]{graphicx,xcolor}
\usepackage{tabularx}
\usepackage{booktabs}
\usepackage{textcomp}
\usepackage{amsmath}
\usepackage{amssymb}
\usepackage[]{graphics}
\usepackage{amssymb}
\usepackage{amsmath}
\usepackage{color}
\usepackage{ifpdf}
\usepackage{lipsum}
\usepackage{siunitx}
\usepackage{braket}
\usepackage{soul}

\graphicspath{{figs/}}
\newcommand{\executeiffilenewer}[3]{%
\ifnum\pdfstrcmp{\pdffilemoddate{#1}}%
{\pdffilemoddate{#2}} > 0 {\immediate\write18{#3}}\fi}

\ifpdf\usepackage{epstopdf}\fi

\begin{document}


\title{Vertical routing of spinning dipoles radiation from a chiral metamembrane}

\author{S.~A.~Dyakov}
\email[]{e-mail: s.dyakov@skoltech.ru}
\affiliation{Skolkovo Institute of Science and Technology, Nobel Street 3, 121205 Moscow, Russia}

\author{I.~M.~Fradkin}
\affiliation{Skolkovo Institute of Science and Technology, Nobel Street 3, 121205 Moscow, Russia}
\affiliation{Moscow Institute of Physics and Technology, Institutskiy pereulok 9, Moscow Region 141701, Russia}

\author{N.~A.~Gippius}
\affiliation{Skolkovo Institute of Science and Technology, Nobel Street 3, 121205 Moscow, Russia}

\author{S.~G.~Tikhodeev}
\affiliation{A.~M.~Prokhorov General Physics Institute, RAS, Vavilova 38, Moscow, Russia}
\affiliation{Faculty of Physics, Lomonosov Moscow State University, 119991 Moscow, Russia}


\date{\today}
\begin{abstract}
 We propose a perfect photonic router based on a specially designed chiral bi-metasurface membrane for spin-polarized point light sources. Due to the mirror symmetry breaking in the chiral metamembrane, the radiation power flux of the clockwise and counterclockwise spinning dipoles to the opposite sides of the slab becomes different. We show that spinning dipoles in the specially designed chiral D$_4$-symmetrical bi-metasurface membrane  can emit light either upwards or downwards depending on their rotation direction. We attribute this phenomenon to the Fano-resonance effect which is a result of the guided modes coupling with the far field. We show the advantage of D$_4$-symmetrical structures for the achievement of 100\% routing efficiency. This phenomenon can find applications in spintronics for spin-selective inter-chip coupling or as a measurement tool of spin polarization in memory cells.


\end{abstract}
\maketitle


\section{Introduction}

Ability to control the polarization-sensitive directivity of light propagation in various photonic structures attracts a great interest of researchers due to new opportunities they open in optoelectronics, quantum information processing, and bio-sensing. It is important for many applications to engineer and enhance a circular dichroism of chiral photonic structures. At nanoscale, the degrees of freedom, represented by the orbital angular momentum of light and its circular polarization, influence each other \cite{bliokh2015spin, Lodahl2017, chong2020generation, abujetas2020spin, wei2020momentum,zambon2019optically, bliokh2019geometric, tsesses2019spin, wozniak2019interaction, wang2019induced, dyakov2018magnetic,dyakov2018circularly,zanotto2019photonic,zanotto2020chiral}. This phenomenon is referred to as spin-orbit interaction of light, it can bring new functionalities to optical nano-devices based on chiral light-matter interaction.

Various photonic systems have recently been exploited for directional coupling of light into different optical modes~\cite{Luxmoore2013, Lin2013, Shitrit2013,lin2013polarization, shitrit2013spin, Kapitanova2014, Petersen2014, Mitsch2014,  Lodahl2017, Spitzer2018, Liu2019, fernandez2019new, feis2020helicity,lei2020enhanced, sinev2020steering}. For example, one can use hyperbolic metamaterials for polarization-controlled routing of emission from spinning dipoles ~\cite{Kapitanova2014, Liu2019}, whereas in Ref.\,\cite{Spitzer2018} routing of exitonic emission in a diluted-magnetic-semiconductor quantum well in hybrid plasmonic semiconductor structures has been demonstrated. As to the quantum information, e. g., in Ref.\,\cite{Luxmoore2013} the authors presented a scheme for interfacing an optically addressed spin qubit to a path-encoded photon using a crossed waveguide device.

\begin{figure}[b!]
\centering
\includegraphics[width=1\linewidth]{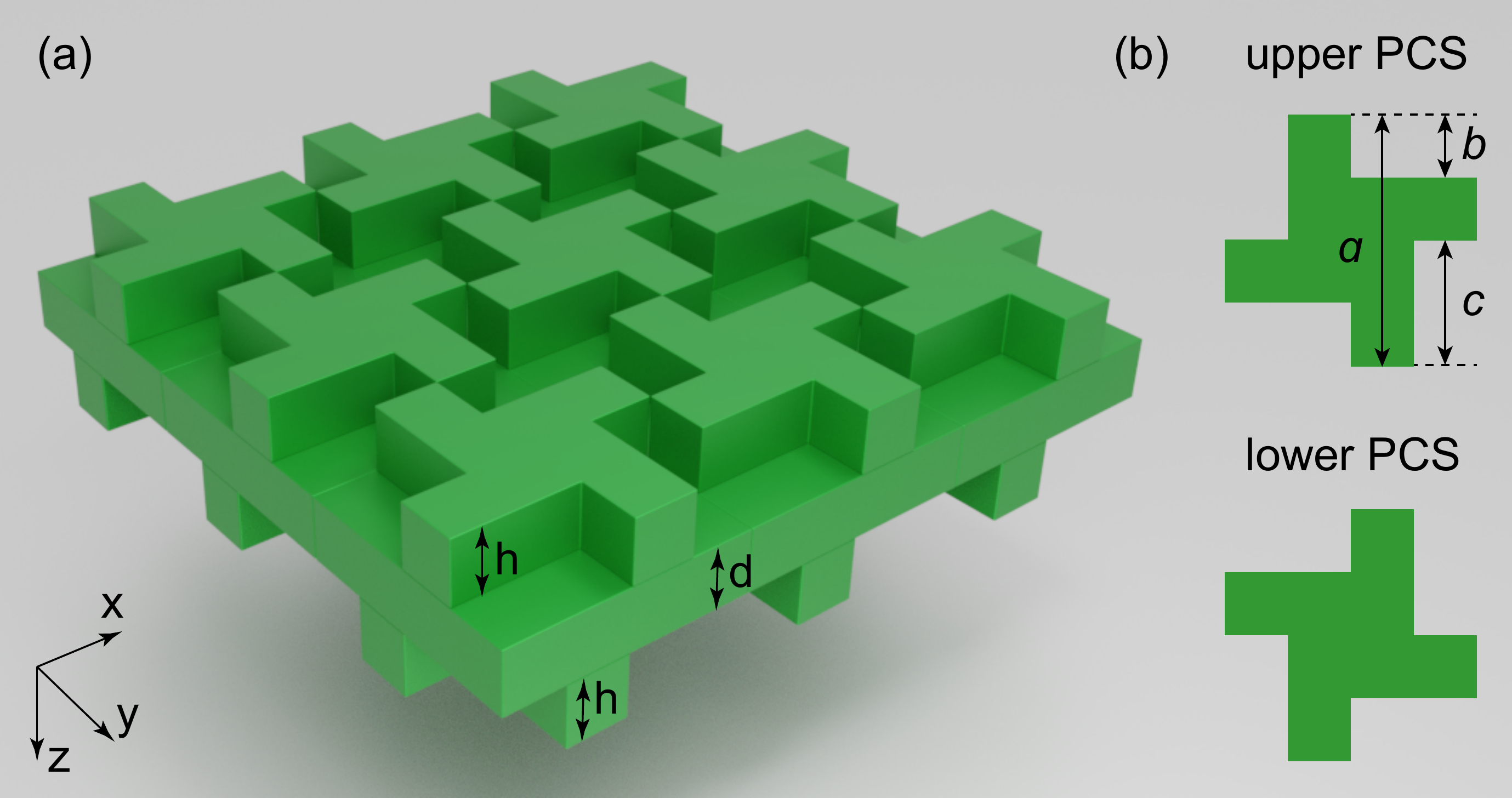}
\caption{(Color online) (a) Schematics of a silicon structure with chiral photonic crystal slabs. The geometrical parameters are $a=801$\,nm, $b=a/4$, $c=a/2$, $h=686$\,nm, $d=401$\,nm. (b) Unit cells of the upper and lower photonic crystal slabs (PCS), as seen from the structure top. }
\label{sample}
\end{figure}

In the above examples, the light routing was realized mostly inside a planar structure assisted by waveguided modes or surface plasmon polariton modes. Subsequent manipulation with these photons will require additional devices, which may impose some limitations on the size and sensitivity of the system. In some applications, e. g., for inter-chip coupling in spintronics, it would be promising to direct the light emission of spin-polarized light sources immediately to the opposite normals of a planar structure.

To demonstrate the possibility of vertical radiative routing of spin-polarized light sources, we consider a periodic silicon membrane with a chiral morphology (see Fig.\,\ref{sample}). It consists of a homogeneous slab of thickness $d$ sandwiched between two photonic crystal slabs of equal thicknesses $h$. The upper and lower photonic crystal slabs are mirror symmetric with respect to the the vertical mirror plane. We will demonstrate that such a D$_4$ symmetrical structure can be a perfect photonic router. The spin-polarized light sources are simulated by spinning point dipoles, i.e., by the classical model of radiating molecules or quantum dots undergoing a spin-polarized transition.


\section{Theoretical method}
In this work, we use the Fourier modal method in the scattering matrix form~\cite{Tikhodeev2002b} (also known as Rigorous Coupled Wave Analysis~\cite{moharam1995formulation}) to calculate the optical properties of a chiral metamembrane. The power flow of dipole's radiation through the horizontal planes at immediately opposite sides of the metamembrane is calculated as a $z$-projection of the Poynting vector integrated over the structure period. In momentum space it is replaced with summation over the Floquet-Fourier harmonics:
\begin{equation}
    P = \frac{c}{16\pi}(\mathrm{E}^\dagger_x \mathrm{H}_y + \mathrm{E}_x \mathrm{H}^\dagger_y - \mathrm{E}^\dagger_y \mathrm{H}_x - \mathrm{E}_y \mathrm{H}^\dagger_x),
\end{equation}
where $\mathrm{E}_{x,y}$ and $\mathrm{H}_{x,y}$ are the vectors of Floquet-Fourier components of $x$- and $y$- projections of electric and magnetic vectors of dipole's radiation field, $c$ is the speed of light and dagger denotes the conjugate transpose. According to the Fourier modal method formalism, $\mathrm{E}_{x,y}$ and $\mathrm{H}_{x,y}$ are found from the vectors of amplitudes by means of the material matrix $\mathbb{F}$:
\begin{equation}
    \begin{bmatrix}
    \mathrm{E}_x\\\mathrm{E}_y\\\mathrm{H}_x\\\mathrm{H}_y
    \end{bmatrix}_\mathrm{u}=
    \mathbb{F}_\mathrm{u}
    \begin{bmatrix}
    \vec{\mathrm{o}} \\ \vec{\mathrm{u}}
    \end{bmatrix}\hspace{30pt}
    \begin{bmatrix}
    \mathrm{E}_x\\\mathrm{E}_y\\\mathrm{H}_x\\\mathrm{H}_y
    \end{bmatrix}_\mathrm{d}=
    \mathbb{F}_\mathrm{d}
    \begin{bmatrix}
    \vec{\mathrm{d}} \\ \vec{\mathrm{o}}
    \end{bmatrix},
\end{equation}
where indices \textit{u} and \textit{d} mean that corresponding quantities are taken in the substrate or superstrate; $\vec{\mathrm{o}}$ is the zero vector of the same size as $\vec{\mathrm{u}}$ and $\vec{\mathrm{d}}$, the outgoing hypervectors which can be found by the method of oscillating currents~\cite{whittaker1999scattering, whittaker2000inhibited, Taniyama2008, lobanov2012emission, fradkin2019fourier}:
\begin{align}
 \vec{\mathrm{u}} &= \mathbb{S}^{\mathrm{u}}_{22}\left(\mathbb{S}^{\mathrm{d}}_{21}\mathbb{S}^{\mathrm{u}}_{12}-\mathbb{I}\right)^{-1}\left(\vec{\mathrm{j}}_\mathrm{d}-\mathbb{S}^{\mathrm{d}}_{21}\vec{\mathrm{j}}_\mathrm{u}\right),   \label{eqn:updn1}\\
   \vec{\mathrm{d}} &= \mathbb{S}^{\mathrm{d}}_{11}\left(\mathbb{I}-\mathbb{S}^{\mathrm{u}}_{12}\mathbb{S}^{\mathrm{d}}_{21}\right)^{-1}\left(\vec{\mathrm{j}}_\mathrm{u}-\mathbb{S}^{\mathrm{u}}_{12}\vec{\mathrm{j}}_\mathrm{d}\right).
 \label{eqn:updn2}
\end{align}
In equations (\ref{eqn:updn1}-\ref{eqn:updn2}) $\mathbb{S}^{\mathrm{u,d}}$ are the upper and lower partial scattering matrices \cite{Tikhodeev2002b, lobanov2012emission} calculated at a given frequency and emission angle; $\vec{\mathrm{j}}_\mathrm{u}$ and $\vec{\mathrm{j}}_\mathrm{d}$ are the hypervectors of oscillating dipole's current that are found from its Floquet-Fourier components $\mathrm{J}_x$, $\mathrm{J}_y$ and $\mathrm{J}_z$ by the use of the material matrix $\mathbb{F}$ of the layer where the dipole is located:
\begin{equation}
    \begin{bmatrix}
    \vec{\mathrm{j}}_\mathrm{u} \\ \vec{\mathrm{j}}_\mathrm{d}
    \end{bmatrix}
    =\mathbb{F}^{-1} \begin{bmatrix}
    -K_x\tilde\varepsilon^{33}\mathrm{J}_z/k_0\\
    -K_y\tilde\varepsilon^{33}\mathrm{J}_z/k_0\\
    -i\mathrm{J}_\mathrm{y}+i\tilde\varepsilon^{23}\mathrm{J}_z\\
    +i\mathrm{J}_\mathrm{x}-i\tilde\varepsilon^{13}\mathrm{J}_z\\
    \end{bmatrix}.
\end{equation}
Here $\tilde\varepsilon$ is the 3$\times$3 block matrix with components that evolve from the Fourier transform of dielectric permittivity tensor \cite{Weiss2009a} calculated in accordance with Li's factorization rules \cite{li1997new}; $K_{\mathrm{x,y}}$ are the diagonal matrices of $x$- and $y$-components of photon quasimomentum vector of different diffraction orders; $k_0$ is the photon wavenumber in vacuum.

To simulate the emission of a rotating point dipole positioned at a spatial coordinate $\vec{r}_0$ in terms of the Fourier modal method, we calculate the harmonics of current $\mathrm{J}_\alpha$ ($\alpha = x,y,z$) as\begin{equation}
    \mathrm{J}_\alpha=j_\alpha e^{-i\vec{r}_0\left(\vec{k}_\parallel+\vec{G}_{mn}\right)},
\end{equation}
where $j_\alpha$ are the components of current in real space, $\vec{k}_\parallel=\left[k_x,k_y\right]$ is the in-plane wavevector, $\vec{G}_{mn} = \left[\frac{2\pi m}{a},\frac{2\pi n}{a}\right]$ is the basis of vectors in reciprocal space, $m$ and $n$ are integers. The current is set as $\vec{j}=\left[1, \pm i, 0\right]$ where the sign "+" (or "-") corresponds to a counterclockwise (or clockwise) rotating dipole moment seen from the positive $z$-direction. We denote the corresponding dipoles as $\sigma^+$ and $\sigma^-$, respectively.

In what follows, we normalize the radiation power flux from a dipole in membrane to the radiation power flux from the same dipole in free space: $I = P/P_0$; we define this ratio as an emissivity.

We use the dielectric permittivity of Si from  Ref.\,\onlinecite{palik1998handbook}.


\section{Results}
\begin{figure}[b!]
\centering
\includegraphics[width=1\linewidth]{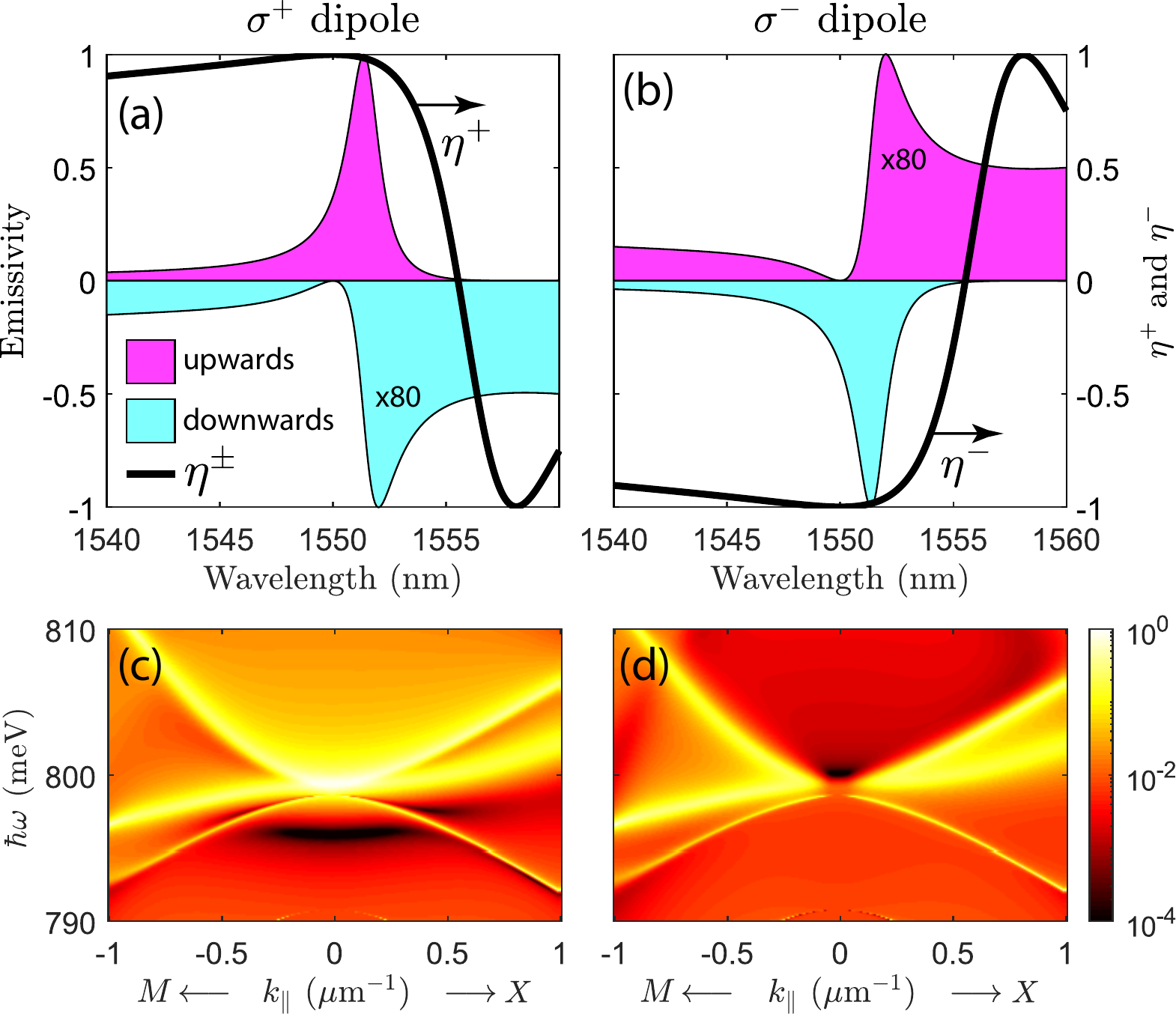}
\caption{(Color online) Normalized emissivity spectra in directions up (magenta area) and down (cyan area) of (a) $\sigma^+$ and (b) $\sigma^-$ spinning dipoles. Black solid lines denote the corresponding routing efficiencies $\eta^\pm$. In-plane wavenumber (in $\Gamma - X$ and $\Gamma - M$ directions) and frequency dependence of the normalized emissivity of the $\sigma^+$ dipole in the vicinity of  $\lambda=$~1.55~${\mu}$m ($\hbar\omega=$~800~meV) in  directions (c) up and (d) down.}
\label{spectra}
\end{figure}

\begin{figure}[t!]
\centering
\includegraphics[width=1\columnwidth]{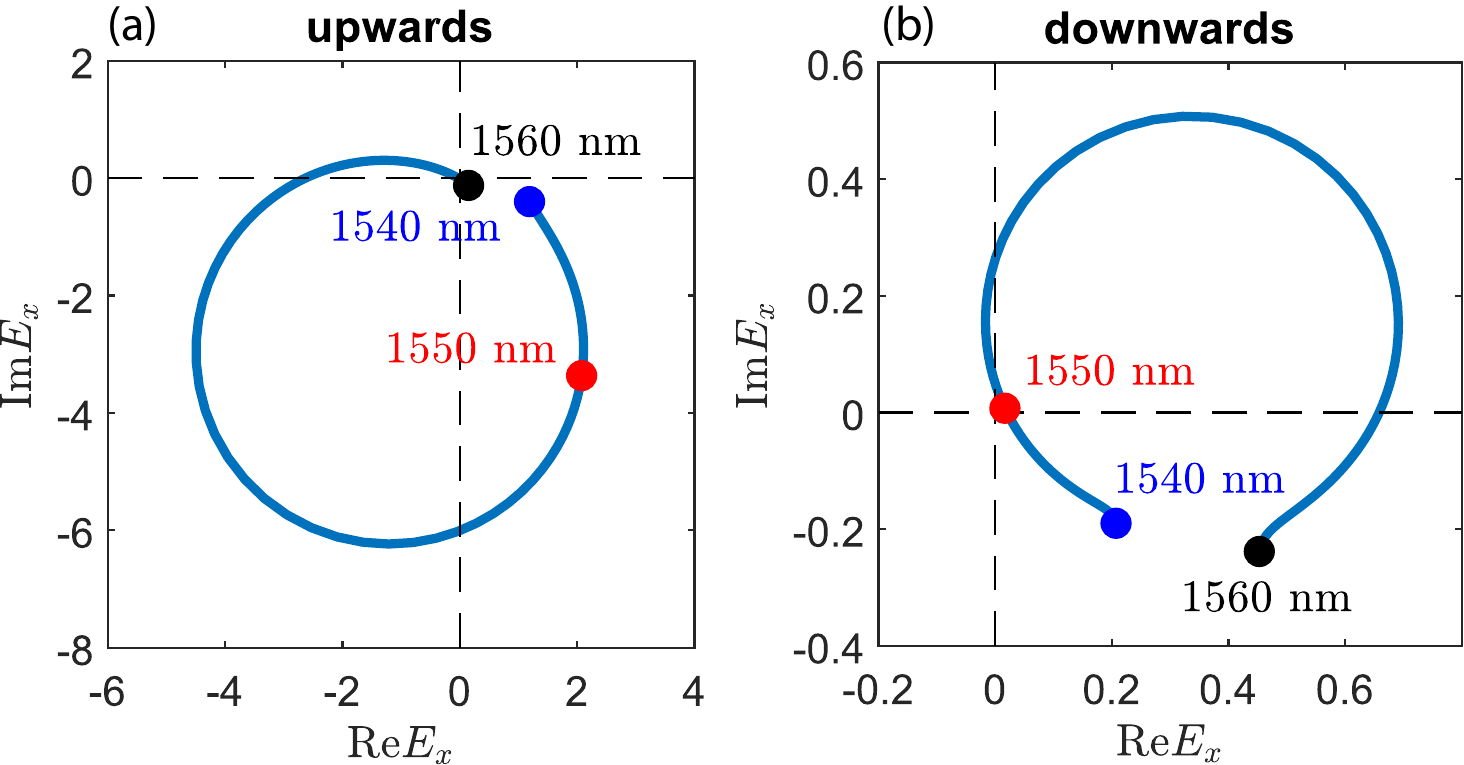}
\caption{(Color online) Complex plane diagrams of $x$-component of the electric field radiated by $\sigma^+$ dipole in directions (a) up and (b) down. Electric field is calculated in the near field of the membrane but only the propagating harmonic is taken into consideration.}
\label{circles}
\end{figure}

\begin{figure*}[t!]
\centering
\includegraphics[width=1\linewidth]{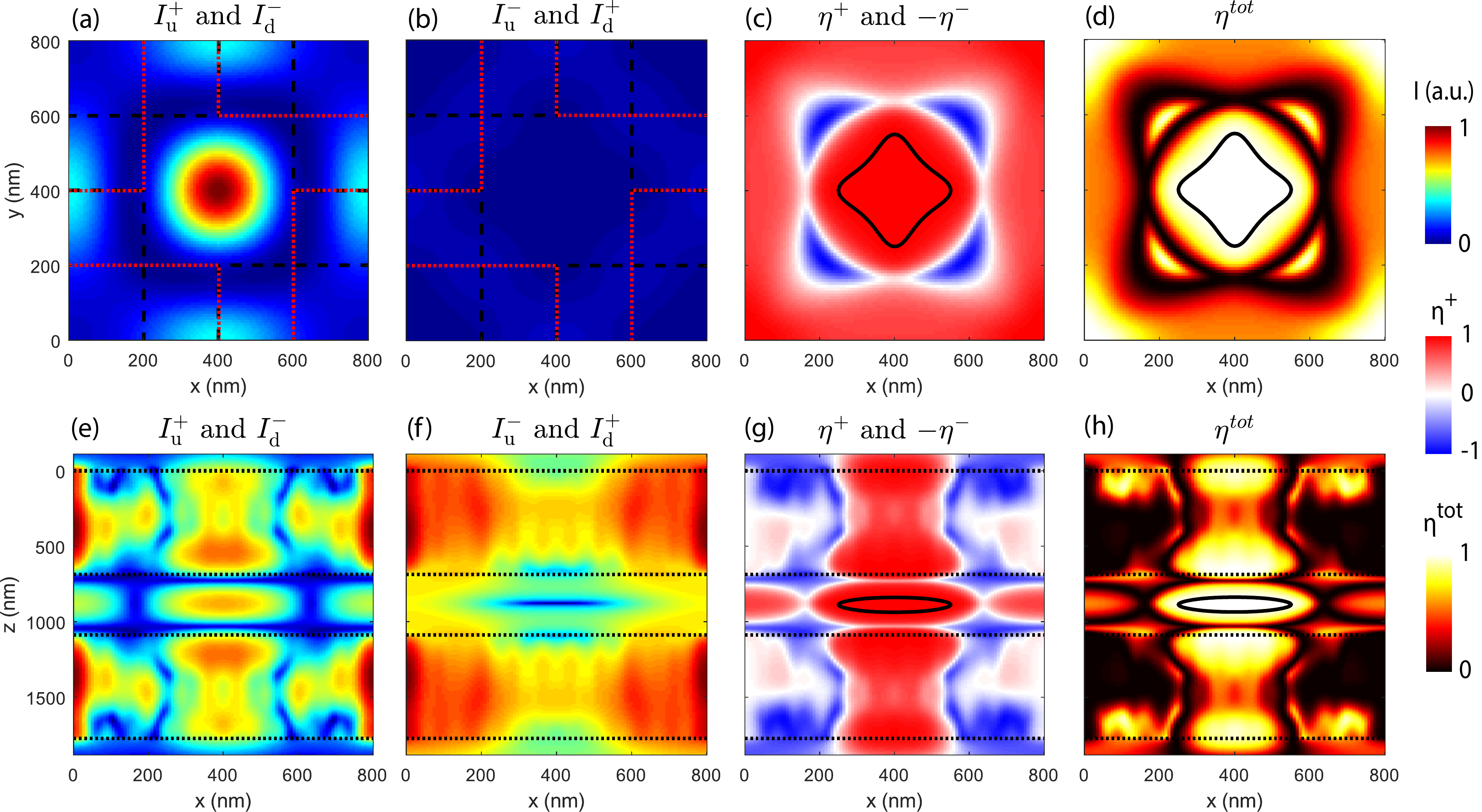}
\caption{(Color online) Dependence of the emissivity (a,d,e,f) in directions up (a,e) and down (b,f) on the $\sigma^+$ dipole's position within an $xy$-plane (a,b) and an $xz$-plane (b,f) which pass through the center of the unit cell. Dependence of the routing efficiency of $\sigma^+$ dipoles (c,g) and the total routing coefficient (d,h) on the dipole position. Black dashed and red dotted lines in (a,b) show the boundaries of pillars in the upper and lower photonic crystal slabs. Black dotted lines in (e--h) show the boundaries between slabs. Black lines in (c,d,g,h) bound the regions where $|\eta^\pm|>0.99$. Color scales on the right denotes emissivity, the routing efficiency, and the total routing coefficient.}
\label{position}
\end{figure*}


The calculated normalized emissivity of $\sigma^+$ and $\sigma^-$ rotating dipoles located in the center of the unit cell of the optimized chiral metamembrane in directions up and down  perpendicular to the structure are shown in Fig.\,\ref{spectra}(a,b). Figures \ref{spectra}(c,d) show the calculated frequency and in-plane wavenumber (in directions $\Gamma - X$ and $\Gamma - M$) dependencies of the normalized emissivity $I^{+,-}_\mathrm{u, d}$ in directions up and down of $\sigma^+$ or $\sigma^-$ rotating dipoles summed over both polarization states of the emitted light. The emissivity of the optimized chiral metamembrane in a wider energy-wavenumber range is shown in Fig.\,S1 of the Supplemental Material(SM) ~\footnote{See Supplemental materials at http://link.aps.org/supplemental/xxx for the emissivity of the optimized chiral metamembrane in a wider energy-wavenumber range.}. Note that all panels of Fig.~\ref{spectra} show the emissivity which is normalized to its maximum which is $\approx 167$. This nearly two orders of magnitude enhancement is due to the Fano resonance effect discussed below. 

Interestingly, the momentum dependence of the emissivity downwards near $\Gamma$-point in Fig.~\ref{spectra}(d) looks very much like a typical behaviour of the symmetry-protected bound state in continuum (BIC), see, e. g., in Ref.~\onlinecite{Sadrieva2019} and references therein. However, here the emission vanishes in the downwards direction only, and for the $\sigma^+$ dipole. Whereas it is open and resonantly enhanced in the opposite  direction, upwards (see in Fig.~\ref{spectra}(c)). Correspondingly, there is no shrinking to zero of the resonance linewidth, only a relative narrowing due to a partial closing of the emission channels. Analogous situation is well known, e. g., for higher resonances in symmetric points of the Brillouin zone when the symmetry protection works only for highly symmetric directions, whereas the diffraction channels remain open~\cite{Tikhodeev2002b}.

One can see in Fig.~\ref{spectra} that for both types of dipoles, the up and down light emission is different. The structure is optimized so that at a vacuum wavelength of  $\lambda=1.55$\,${\mu}$m the downwards emissivity of the $\sigma^+$ dipole is zero while for upwards direction it is close to the maximum. To describe this asymmetry quantitatively we introduce the routing efficiencies $\eta^+$ and $\eta^-$, as well as the total routing efficiency $\eta^{\mathrm{tot}}$ defined as
\begin{equation}
    \eta^\pm=\frac{I_\mathrm{u}^\pm-I_\mathrm{d}^\pm}{I_\mathrm{u}^\pm+I_\mathrm{d}^\pm}\\,\hspace{30pt}
    \eta^{\mathrm{tot}} = -\eta^+\eta^- .
\end{equation}
It becomes possible for the optimized metamembrane to reach $\eta^+=1$, $\eta^-=-1$, $\eta^{\mathrm{tot}} = 1$ at $\lambda=1.55$\,${\mu}$m. Such a perfect routing is due to the vanishing emissivities $I_{\mathrm{d}}^+$ and $I_{\mathrm{u}}^-$ at this wavelength which is attributed to the Fano effect. Peaks in Fig.\,\ref{spectra} have asymmetric Fano-type shapes; they represent the quasiguided modes (also known as quasinormal guided modes \cite{Leung1994,Tikhodeev2002b,Gippius2005c,Muljarov2010,Kristensen2014,Alpeggiani2017,Lassalle2018,Lalanne2019,gras2019quasinormal}) that appear in the emission spectra due to the grating assisted coupling of the emitted light with photon continuum of the far field~\cite{Fano1941,Fano1961,luk2010fano,Tikhodeev2002b,PhysRevB.93.205413,dyakov2012surface,dyakov2018plasmon}.

\begin{figure*}[t!]
\centering
\includegraphics[width=0.9\linewidth]{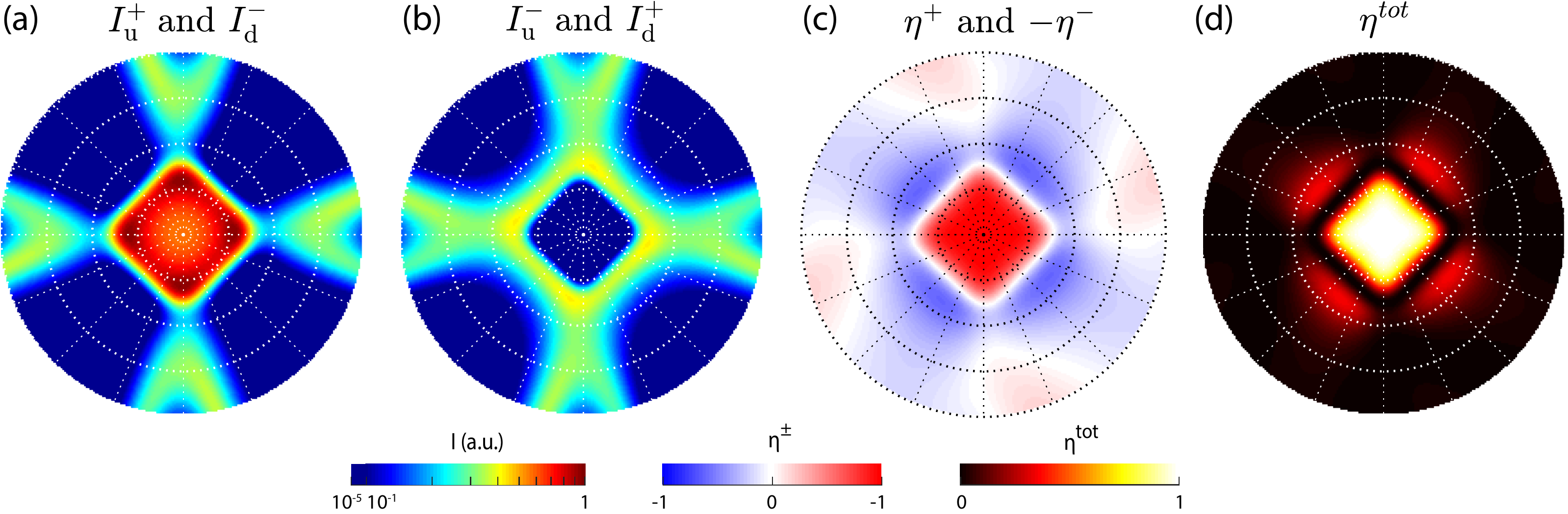}
\caption{(Color online) Angular emission diagrams within $\pm 10 ^\circ$ cone angle of $\sigma^+$ dipoles (a) in directions up and (b) down, (c) the routing coefficient $\eta^+$ and (d) the total routing efficiency. Colorscale is shown in the bottom.}
\label{angular}
\end{figure*}

In general, the asymmetric Fano shape appears in reflection, transmission or emissivity spectra of a photonic crystal slab as a result of interference of its eigenmode $e(\omega)$ with a smooth background term $b(\omega)$ \cite{Gippius2005c}:
\begin{equation}
    x(\omega)=e(\omega)+b(\omega) = \frac{f}{\omega - \omega_0 + i\gamma} +b(\omega),
    \label{eqFano}
\end{equation}
where $f$ is the oscillator strength, $\omega_0$ is the eigenfrequency and $\gamma$ is the damping coefficient. In the case of a pure resonance ($b=0$), the complex plane trajectory of the parameter $x(\omega)$ is represented by a circle arond zero and the emissivity spectrum of $|x(\omega)|^2$ is a symmetric Lorentzian curve. When $b\ne0$ the spectrum $|x(\omega)|^2$ becomes asymmetric and at certain geometrical parameters and frequency can be zero. When $b(\omega)$ is a weak function of $\omega$, Eqn.\,(\ref{eqFano}) is often interpreted as an interference of a discrete state $e(\omega)$ with a continuum of states $b(\omega)$ \cite{luk2010fano}.

Complex plane diagrams of the $x$-component of the electric field radiated by $\sigma^+$ dipole in directions up and down are shown in Fig.\,\ref{circles}. One can see that the relative phase of $E_x$ changes by a value of $2\pi$ with an increase of $\lambda$ from 1540 to 1560\,nm. At $\lambda\approx$\,1550\,nm for downwards emission the phase trajectory of $E_x$ passes through the origin while for upwards emission the zero point on the phase trajectory is relatively far from it. This results in $\eta^{\mathrm{tot}}=1$ at this wavelength. For $\lambda=$\,1553.6\,nm the situation is opposite and $\eta^{\mathrm{tot}}=1$ too. Similar diagrams can be obtained for the $E_y$.



For the practical realization, it is interesting to study the stability of this effect relative to the dipole position. The dependencies of the emissivity in directions up and down on the $\sigma^+$ dipole's position within the horizontal and vertical planes which pass through the center of the cell are shown in Fig.\,\ref{position}. The corresponding dependencies for the $\sigma^-$ dipoles are the same (not shown), however, one must swap the directions up and down. One can see that these dependencies have maximum and minimum in the central region of the cell. In this region the emission of $\sigma^+$ (or $\sigma^-$) dipole is not coming out in downwards (or upwards) direction and the corresponding routing efficiency is $\eta^+=1$ (or $\eta^-=-1$). There are also the local minima from where the emission in directions up or down is sufficiently suppressed, however, it is not strictly zero, so that $|\eta^\pm|<1$. The resulting dependencies of the routing efficiencies $\eta^\pm$ on the dipole position consist of regions with positive and negative values (Fig.\,\ref{position}(c,g)). Figure \,\ref{position}(d,h) shows the total routing efficiency, solid lines bound the regions with $|\eta^{\mathrm{\pm}}|>0.99$. One can see that efficient routing is possible for dipoles located in rather a large volume of the unit cell. If a quantum dot undergoing the spin-polarized transition is placed within this volume, its upwards or downwards emission will be strongly suppressed.

The angular emission diagrams of $\sigma^+$ dipoles within $\pm$10$^\circ$ cone are shown in Fig.\,\ref{angular}. The lines of high emissivity represent the quasiguided modes of the chiral metamembrane. One can see from Fig.\,\ref{angular} that the effect of routing with $\eta>0.99$ exists only in the $\sim \pm2.5^\circ$ cone around the normal angle.

\section{Discussion}
In this work as a proof of principle of routing the emission of spinning dipoles, we use the structure with D$_4$ rotational symmetry (Fig.\,\ref{sample}). Let us now demonstrate that for the effect of a perfect routing the D$_4$ symmetry is more suitable than C$_4$ which is often used to obtain circular dichroism. The structure in Fig.\,\ref{sample} would have the C$_4$ symmetry if it was on the substrate or if it had only one of the chiral photonic crystal slabs.

We start our discussion with C$_4$ symmetrical structures as it is a more general case. We consider the electric field generated by an oscillating dipole located in the middle of the cell in the $xy$-plane at an arbitrary $z$-coordinate. The electric field is considered below and above the membrane. We denote the complex-valued electric field components produced by the $x$-polarized dipole moment as
\begin{equation}
    \vec{E}_{\mathrm{u}}^x = [\alpha, \beta]\\,\hspace{30pt}
    \vec{E}_{\mathrm{d}}^x = [\rho, \tau].
\end{equation}
For the C$_4$ symmetrical structures, the fields generated by the $y$-polarized dipole moment have the form:
\begin{equation}
        \vec{E}_{\mathrm{u}}^y = [-\beta, \alpha]\\,\hspace{30pt} \vec{E}_{\mathrm{d}}^y = [-\tau, \rho].
\end{equation}

By the superposition principle, the electric fields from the $\sigma^+$ and $\sigma^-$ polarized dipoles are found as
\begin{align}
    &\vec{E}_{\mathrm{u}}^{\sigma^\pm} = \vec{E}_{\mathrm{u}}^x\pm i\vec{E}_{\mathrm{u}}^y =
    (\alpha\mp i\beta)\left[1,\pm i\right]\nonumber\\
    &\vec{E}_{\mathrm{d}}^{\sigma^\pm} = \vec{E}_{\mathrm{d}}^x\pm i\vec{E}_{\mathrm{d}}^y =
     (\rho\mp i\tau)\left[1, \pm i\right].
\end{align}
With the simultaneous fulfillment of the conditions
\begin{equation}
   \alpha=-i\beta, \qquad \rho=i\tau
   \label{cond}
\end{equation}
 the emission intensities from $\sigma^+$ dipoles in downwards direction and $\sigma^-$ dipoles in upwards direction are zeros and the resulting total routing coefficient $\eta^{\mathrm{tot}}=1$. The electric field components $\alpha$, $\beta$, $\rho$, $\tau$ depend on geometrical parameters of the metamembrane and it can be problematic to simultaneously fulfill the above two conditions in structures with C$_4$ symmetry. Whereas the D$_4$ symmetrical structures, such as shown in Fig.\,\ref{sample}, have one more symmetry operation under which the structure is invariant, namely the rotation by 180$^\circ$ about the $y$-axis (or equivalent $x$-axis). It leads to the fact that a dipole, located in the center of the cell of any D$_4$ symmetrical structure, generates electric field such that $\rho=\alpha$ and $\tau=-\beta$. It makes the conditions (\ref{cond}) equivalent to each other. Thus, for D$_4$ symmetrical structures the condition for the perfect routing is expressed only by one equality $\alpha=-i\beta$ which is much easier to satisfy than both of equalities (\ref{cond}) simultaneously. See Fig.\,S2 in SM at \footnote{See Supplemental Material at http://link.aps.org/supplemental/xxx for dependencies of the total routing coefficient on the geometrical parameters of the chiral metamembrane.} for dependencies of $\eta^{tot}$ on the geometrical parameters. Using this condition we obtain that in D$_4$ symmetrical structures the emission from $\sigma^+$ and $\sigma^-$ dipoles is circularly polarized:
\begin{align}
    &\vec{E}_{\mathrm{u}}^{\sigma^+} = [2\alpha, 2i\alpha],   &  \vec{E}_{\mathrm{u}}^{\sigma^-} &= [0, 0],\nonumber \\
    &\vec{E}_{\mathrm{d}}^{\sigma^+} = [0, 0],&  \vec{E}_{\mathrm{d}}^{\sigma^-} &= [2\alpha, -2i\alpha].
    \label{helicity}
\end{align}

From (\ref{helicity}) it follows that the upwards emission from $\sigma^+$ dipole and the downwards emission from $\sigma^-$ dipole both have the same helicity such that an observer looking in the direction from which the light is coming, sees the electric vector describing the circle in the counterclockwise sense. By the simultaneous change of the sign of the right hand side in Eqn.\,\ref{cond}, one can obtain the condition for the opposite helicity of radiated light: $\alpha=i\beta$. In the described chiral metamembrane, the conditions $\alpha=\mp i\beta$ are fulfilled at $\lambda=1.55$\,$\mu$m and $\lambda=1.5536$\,$\mu$m, respectively. It is noteworthy that the directions of electric vector rotation and dipole's rotation are the same. 

\section{Conclusion}
In conclusion, we have theoretically demonstrated the chiral metamembrane with the highest possible circular dichroism of the light emission of clockwise and counterclockwise rotating electric dipoles. In the optimized for optical communication wavelength 1.55~${\mu}$m structure, the $\sigma^+$ dipole radiates light entirely upwards while the $\sigma^-$ dipole radiates entirely downwards. We attribute this effect to the appearance of Fano resonance which occurs due to the grating assisted coupling of guided modes with far-field. We have shown the advantage of D$_4$ symmetric structure for the achievement of this effect. This phenomenon can find application in spintronics as a spin-selective photonic router.

\section{Acknowledgments}
This work was supported by the Russian Science Foundation (Grant \textnumero 16-12-10538$\Pi$).

%


\end{document}